\documentclass[conference]{IEEEtran}
\IEEEoverridecommandlockouts
\usepackage{cite}
\usepackage{amsmath,amssymb,amsfonts}
\usepackage{algorithmic}
\usepackage{graphicx}
\usepackage{textcomp}
\usepackage{subcaption}
\usepackage{xcolor}
\makeatletter
\newcommand{\pcite}[1]{%
  \begingroup\let\cite@adjust\@empty
  ({\cite{#1}})%
  \endgroup
}
\def\BibTeX{{\rm B\kern-.05em{\sc i\kern-.025em b}\kern-.08em
    T\kern-.1667em\lower.7ex\hbox{E}\kern-.125emX}}
\begin{document}

\title{\vspace{6mm}The Effects of Driver Coupling and Automation Impedance on Emergency Steering Interventions% Emergency Situations and Automation Faults%intended and faulty automatic steering interventions
{}
\thanks{Toyota Research Institute (TRI) provided funds to assist the authors with their research but this article solely reflects the opinions and conclusions of its authors and not TRI or any other Toyota entity.}
}

\author{\IEEEauthorblockN{Akshay Bhardwaj, Yidu Lu}
\IEEEauthorblockA{
\textit{University of Michigan Ann Arbor}\\
Michigan, United States \\
akshaybh@umich.edu, luyd@umich.edu}
\and
\IEEEauthorblockN{Selina Pan}
\IEEEauthorblockA{
\textit{Toyota Research Institute}\\
California, United States \\
selina.pan@tri.global}
\and
\IEEEauthorblockN{Nadine Sarter, Brent Gillespie}
\IEEEauthorblockA{
\textit{University of Michigan Ann Arbor}\\
Michigan, United States \\
sarter@umich.edu, brentg@umich.edu}
}

\maketitle
\begin{abstract}
Automatic emergency steering maneuvers can be used to avoid more obstacles than emergency braking alone. While a steer-by-wire system can decouple the driver who might act as a disturbance during the emergency steering maneuver, the alternative in which the steering wheel remains coupled can enable the driver to cover for automation faults and conform to regulations that require the driver to retain control authority. In this paper we present results from a driving simulator study with 48 participants in which we tested the performance of three emergency steering intervention schemes. In the first scheme, the driver was decoupled and the automation system had full control over the vehicle. In the second and third schemes, the driver was coupled and the automation system was either given a high impedance or a low impedance. Two types of unexpected automation faults were also simulated. Results showed that a high impedance automation system results in significantly fewer collisions during intended steering interventions but significantly higher collisions during automation faults when compared to a low impedance automation system. Moreover, decoupling the driver did not seem to significantly influence the time required to hand back control to the driver. When coupled, drivers were able to cover for a faulty automation system and avoid obstacles to a certain degree, though differences by condition were significant for only one type of automation fault.   
\end{abstract}

\section{Introduction}

Progress in the development of vehicle automation and advanced driver assist systems has made driving more efficient, comfortable, and safe \cite{mulder2012sharing, young2007driving}. To date, most vehicle automation is aimed at offloading certain routine tasks from the human driver in non-emergency situations \cite{bhardwaj2020s}. Yet automation can also mitigate or avoid accidents in emergency situations, where fast reaction times and the ability to push control up to vehicle traction limits enable automation to outperform the human driver   \cite{heesen2014interaction, sieber2015automatic, liu2019improving}. An example is emergency collision avoidance that uses automatic braking to prevent or mitigate a collision \cite{klauer2005driver, hesse2013interaction}. 
Above a certain speed and below a certain time-to-collision (shorter than about two seconds), steering maneuvers can result in more successful collision avoidance than braking \cite{heesen2014interaction, adams1994review, sieber2015automatic}.
However, unlike automatic braking systems, automatic steering intervention systems are not yet available on the road \cite{heesen2014interaction, hesse2013interaction, fricke2015driver}. 
Accordingly, development of automatic steering intervention systems has attracted significant interest in the recent past \cite{heesen2014interaction, sieber2015automatic, hesse2013interaction, ruger2015automatic, fricke2015driver}.

To take over control during an emergency situation, the steering wheel can be decoupled from the driver with the use of a steer-by-wire-system \cite{heesen2014interaction, flemisch2010shared}. %\todo[inline, backgroundcolor=blue!10!white,bordercolor=white]{[SP] Consider “To take over control during an emergency situation, the steering wheel can be decoupled from the driver with the use of a steer-by-wire-system”}
However, decoupling the driver invokes liability issues, especially when there is a false activation of the automation system. A decoupled steering wheel during a false activation might render the driver incapable of avoiding a collision that could be prevented %Responsibility for a collision that takes place during a false activation and when the driver is decoupled from the tires must be carried by the automation system, as the argument would remain that the human driver could have prevented the collision if they had been afforded any control 
\cite{heesen2014interaction, ruger2015automatic}. 
Further, the present legal system and code of industrial practice, as well as the principles of human-centered design dictate that a driver should always maintain some degree of control over the vehicle \cite{commitee1968convention,schwarz2006response, young2007driving}.

The issues with decoupling the driver are compounded during automation dropouts (when the automation fails to detect and avoid an obstacle). The literature provides ample evidence that highly automated systems may reduce driver vigilance and situation awareness due to a reduced involvement of the driver in the driving task and a lack of timely feedback of automation actions \cite{endsley1995out, sarter1995world, parasuraman2000model}. A driver who expects they are decoupled from the steering system might become over-reliant on the automation and might fail to prevent a collision during automation dropouts \cite{mulder2012sharing, young2007driving}. %A loss of vigilance may also increase the time required for handing authority back to the human driver, 
Moreover, the time required for the driver to resume control of the vehicle may also increase, leading to additional safety issues \cite{fricke2015driver}.

The alternative is to use Haptic Shared Control  \cite{abbink2012haptic} where the driver remains coupled while the automation system intervenes through a motor on the steering system. However, with a coupled steering wheel, the steering torque applied by the driver may become a disturbance to the automation. %\cite{sieber2015automatic, fricke2015driver, heesen2014interaction}. 
During emergency situations in particular, the startled drivers may fight any steering intervention which can further subvert automation's efforts to prevent accidents \cite{seto2004research, heesen2014interaction, sieber2015automatic, fricke2015driver}.  %Emergency situations are followed by a psychological shock that results in a sudden increase in driver's muscular impedance
%. Such a driver reaction can further subvert the automation's efforts to prevent an accident \cite{sieber2015automatic, fricke2015driver}. 

Is it possible to %choose a mechanical impedance for the automation in coupled driving that would 
suppress driver disturbance yet retain a certain driver authority?  Naturally, the human driver can modulate their driving point impedance (and hence their driving authority) through muscle action and can attempt to overpower the automation in emergency scenarios \cite{mulder2012sharing, petermeijer2015should}. %A high impedance automation system has the potential to reduce the effects of conflicting driver inputs during emergency situations while not fully taking away driver authority. 
% drivers would "prefer" to be coupled. 
But a tradeoff appears to exist between collision avoidance provided by the automation and fault tolerance provided by the driver. Setting the automation impedance high might reduce the effect of driver disturbance at the steering wheel but might also result in an `authoritarian' automation system against which the driver has to compete for control (especially during automation faults when the driver \emph{should} have authority) \cite{petermeijer2015should, mars2014analysis, zwaan2019haptic}. High impedance automation may also risk injuring the driver's hands, especially when acting unexpectedly \cite{fricke2015driver}. 
%While a low impedance automation system may be preferred by drivers and may be  more helpful in avoiding obstacles during automation faults, it may suffer from poor obstacle avoidance performance due to involuntary driver intervention.
%
% how to use the following citations? 
%\cite{petermeijer2015should, young2007driving, kerschbaum2014highly, hesse2013interaction}.

A small number of studies have analyzed driving performance with high and low impedance automation systems and coupled and decoupled steering wheels. However, studies on the effect of high and low levels of automation impedance are limited to lane keeping applications \cite{petermeijer2015should, mars2014analysis, zwaan2019haptic, mulder2012sharing}. 
On the other hand, studies focused on automatic emergency steering interventions are limited to investigations of the effects of haptic and auditory warnings in driving with a decoupled steering wheel \cite{sieber2015automatic, hesse2013interaction}. %or the effect of steering wheel design in driving with decoupled steering wheel \cite{kerschbaum2014highly}. 
Only one study on automatic emergency steering interventions compared the performance of driving with an intermittently decoupled steering wheel to driving with a low impedance coupled steering wheel  \cite{heesen2014interaction}. %However, as authors pointed out, in this study the automation impedance in the coupled steering wheel case was set too low and the driver disturbance resulted in a collision with every obstacle placed on the test track. % [am I right?] result participants driving with coupled steering could not avoid any obstacles in emergency situations. 
%However, in this study, as authors pointed out, the automation impedance was set too low to suppress driver disturbance and avoid any obstacle. 
However, none of the previous studies have focused on understanding the influence of high and low automation impedance in the context of driver coupling during emergency steering interventions.

% However, in this study, all the obstacles were hit in the coupled steering wheel case had a very low impedance

% every obstacle was hit 

% However, in this study all the obstacles were hit in the coupled steering wheel case because the automation impedance was set too low to suppress driver disturbance.

%
%  Agent is automation? Comment is on...? 
%
%The results showed that drivers with a decoupled steering wheel almost always avoided the obstacles whereas those with a coupled steering wheel could never avoid the obstacle. The authors pointed out that the automation impedance was low and was easily disturbed by driver input. 

In the present driving simulator study, we investigated the effect of driver coupling and automation impedance on automatic emergency steering interventions. We designed a decoupled steering scheme and two coupled steering schemes (one with low and the other with high automation impedance), and analyzed the human-automation team performance at avoiding obstacles during emergency situations and automation faults. For half of the participants, we simulated faults by making the automation inactive near an obstacle and for the other half 
by simulating a false automation activation. We also tested the effect of the three coupling schemes on the time needed by the driver to take back control from the automation system.

\section{Methods}

\subsection{Participants}

Forty-eight participants (27 male, 21 female) participated in this study. The participants were between 20 and 30 years old (mean 23.5 years, SD = 3.6 years), had more than two years of driving experience (mean 5.9 years, SD = 3.2 years), and self-reported as having normal or corrected-to-normal vision and normal hearing. All participants provided written informed consent in accordance with a protocol approved by the University of Michigan Institutional Review Board (ID: HUM00164233). 
A complete experiment (including testing, training, and survey) was two hours long. Each participant was provided a financial compensation of \$30 for completing the experiment.

\subsection{Apparatus}

\begin{figure}[h!]
\centering
\begin{subfigure}[h]{0.38\textwidth}
\includegraphics[width=1\textwidth]{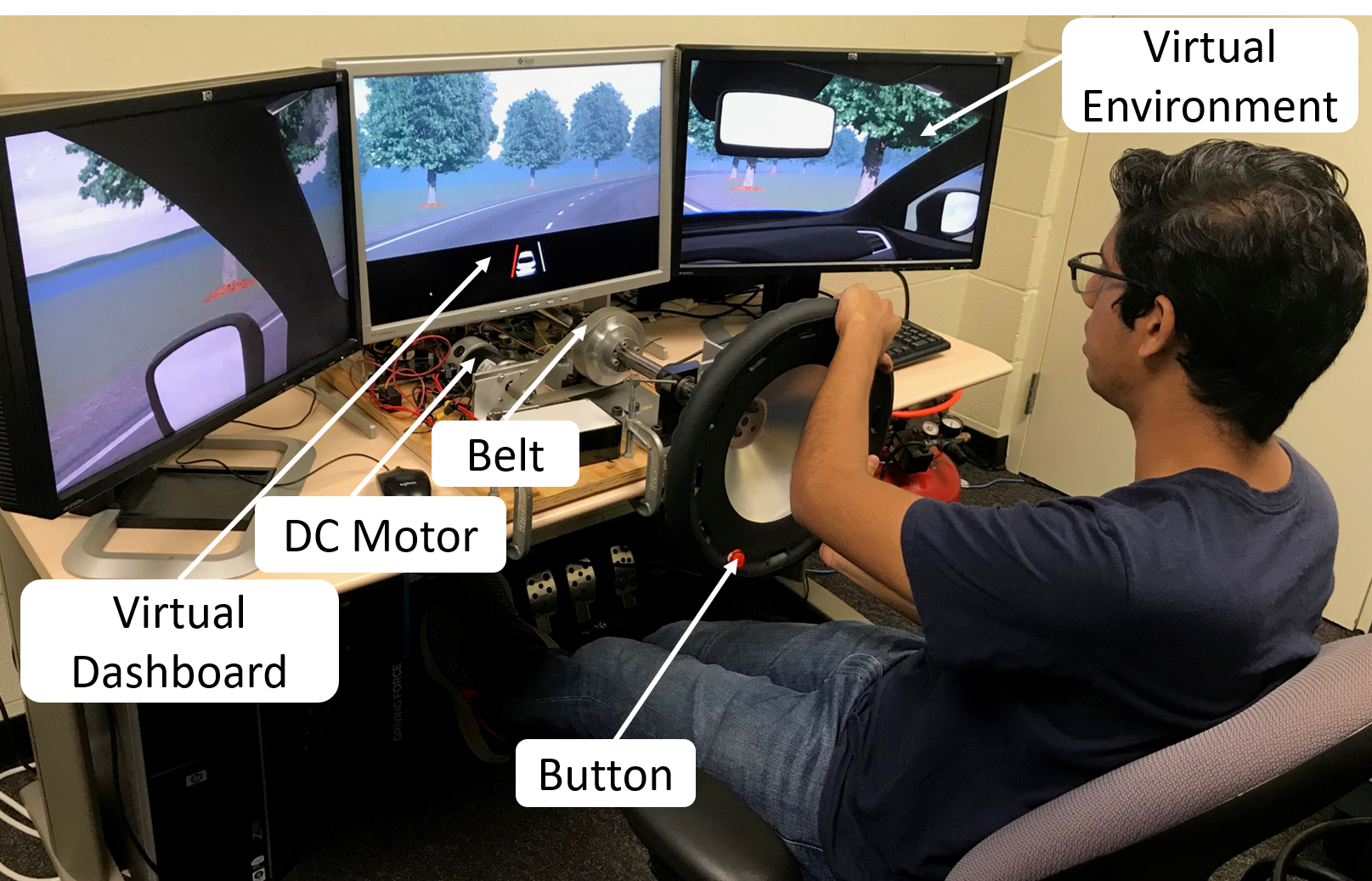}
   \caption{}
   \label{1a}
\end{subfigure}
\par\medskip
\begin{subfigure}[h]{0.38\textwidth}
  \includegraphics[width=1\linewidth]{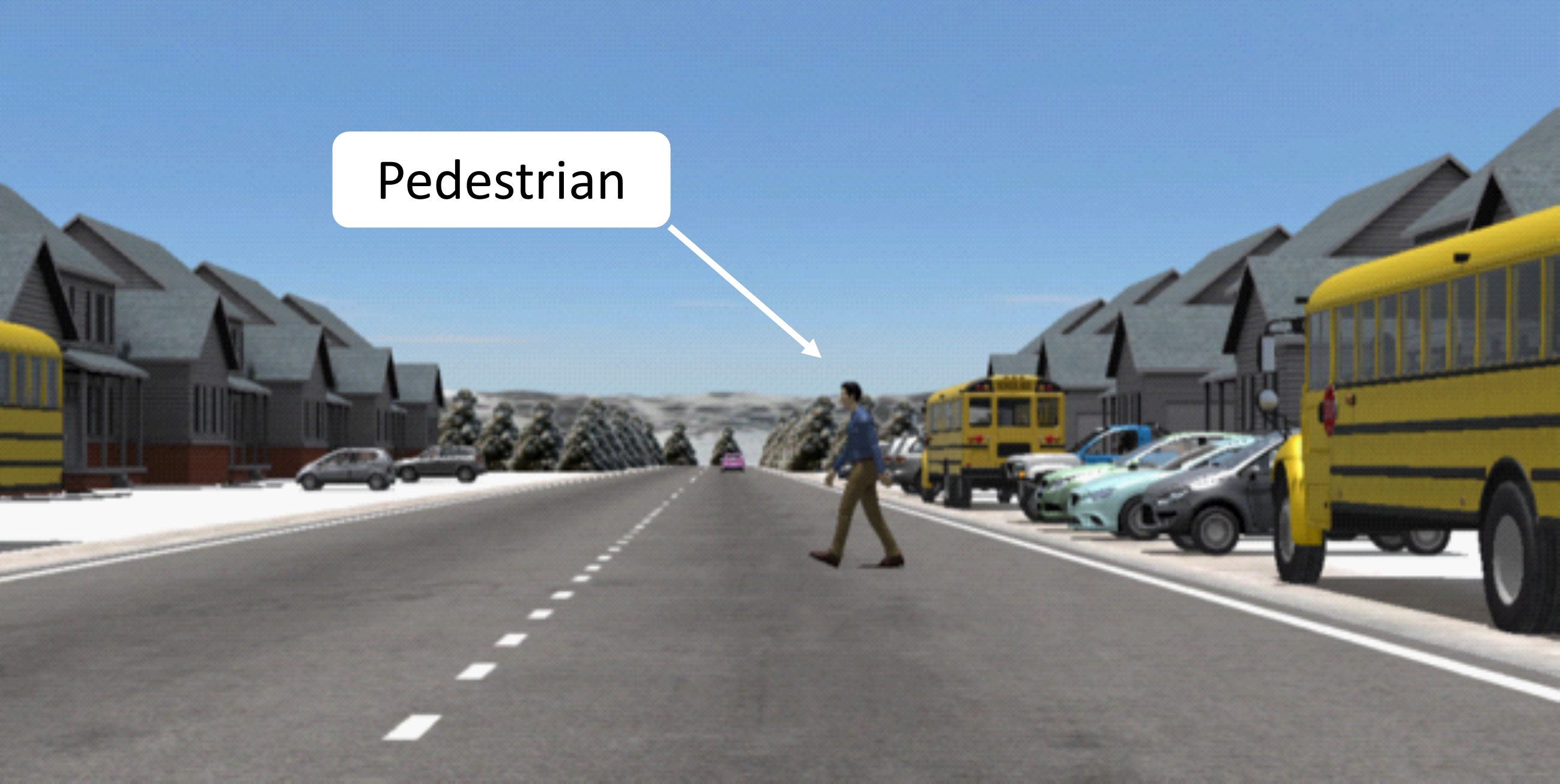}
  \caption{}
  \label{1c}
\end{subfigure}
\par\medskip
\begin{subfigure}[h]{0.35\textwidth}
   \includegraphics[width=1\linewidth]{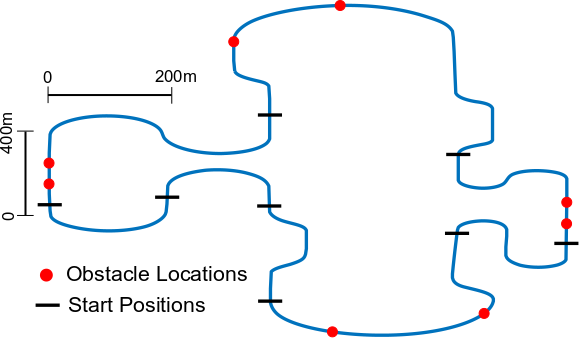}
   \caption{}
   \label{1b} 
\end{subfigure}
\par\medskip
\begin{subfigure}[h]{0.48\textwidth}
\includegraphics[width=1\textwidth]{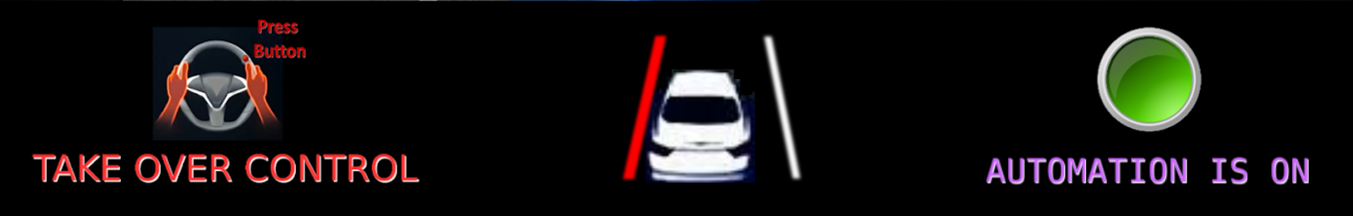}
   \caption{}
   \label{1cc} 
\end{subfigure}

\caption[]{Experimental setup. (a) A subject performing the test on the fixed-base driving simulator. (b) CarSim virtual environment depicting the scenario in which a pedestrian unexpectedly enters the road. (c) Top view of the driving track (navigated clockwise) indicating the obstacle locations and starting positions. (d) Virtual dashboard in the virtual environment showing warnings and notifications.}
\end{figure}

% The experimental apparatus was a custom fixed-base driving simulator featuring a motorized steering wheel (Fig. \ref{1a}). A DC motor (AmpFlow A28-150, Belmont, CA) was coupled to the steering wheel (Speedway 38 cm solid aluminum wheel, Lincoln, NE) through a timing belt with a 72:15 mechanical advantage, making up to 66 N-m torque available to be imposed on the human driver. A 10,000 count per revolution optical encoder (US Digital HB6M, Vancouver, WA) was attached to the steering shaft and the motor was equipped with a 2048 count per revolution optical encoder (US Digital HB6M). In addition, the steering wheel was equipped with a red button within easy reach of a participant’s thumb on the steering wheel to take back control from automation. The virtual driving environment was displayed on three 24-inch LCD widescreen monitors positioned about 140 cm from the participant.

% The experimental apparatus was a custom fixed-base driving simulator featuring a motorized steering wheel (Fig. \ref{1a}). Hardware details of the fixed-base simulator are described in \cite{bhardwaj2020s}. However, one change has been made to the simulator: the virtual driving environment was displayed on three 24-inch LCD widescreen monitors instead of one 50 cm widescreen monitor. 

% Details pertinent to automation motor and steering wheel can be found in 

The experimental apparatus was a custom fixed-base driving simulator featuring a motorized steering wheel (Fig. \ref{1a}). Details pertinent to the steering wheel design, automation motor, encoders, and their assembly can be found in \cite{bhardwaj2020s}. The virtual driving environment was displayed on three 24-inch LCD widescreen monitors positioned about 140 cm from the participant. The vehicle dynamics and control and the virtual environment were implemented in CarSim (Mechanical Simulation Corporation, Ann Arbor, MI) and Simulink (Mathworks, Natick MA) and were computed in real-time on a Dell Precision 5820 Tower Workstation computer using an Intel Xeon W-2125 Quad-Core processor. CarSim math models and Simulink code were computed at 1000 Hz and the graphical display was rendered at 50 Hz. 

The virtual environment was created in CarSim VS Visualizer, and appeared as shown in Fig. \ref{1c}. It featured a D-Class Sedan vehicle and a two-way road with various landmarks and vehicles that provided motion cues during driving. The vehicle traveled at a constant speed of 60 km/h using `Constant Target Speed' control in CarSim. Neither the participant nor the automation system had any control over speed. The two-way road was 8 m wide with 4 m wide lanes and a dashed line separated traffic in two directions. The track width of the vehicle was about 2.1 m. The entire road was 6 km long and the obstacle locations and starting stations on the road were randomized as shown in Fig. \ref{1b}. Visual notifications and warnings were provided to the participants through a virtual dashboard on the central monitor as shown in Fig. \ref{1cc}.  %\todo[inline, backgroundcolor=blue!10!white,bordercolor=white]{[SP] Consider swapping the order of pictures c and d to match the order in which you describe them here.  I agree with the order in which you described them here.  If you do switch them, be sure to update all instances where 1c and 1d are referenced in the rest of the paper.}
 Finally, audio alerts were provided to the participants through a speaker located on the right side of the steering wheel. The visual and audio alerts are further described in section \ref{procedure}.

% A `Take Over Control' notification appeared when automation brought the vehicle back to the center of the right lane after avoiding the obstacle. Four seconds after the first appearance of the `Take Over Control' notification, monotone auditory alerts generating one ``beep" per second were sent from a speaker to remind the driver to take over. The participants were instructed to keep the vehicle in the center of the right lane. A lane departure warning appeared when the deviation of the vehicle from the center of the right lane exceeded 0.4 m. A `AUTOMATION Is On' notification appeared when the automation turned on to avoid the obstacle.

%left tires of the vehicle were within 0.4 m of the centerline or when the right tires were within 0.4 m of the right lane marking

%Auditory Alert

\subsection{Automation Design and Haptic Feedback}

We used a pure pursuit controller to develop an automation system capable of lane keeping and obstacle avoidance. A pure pursuit controller generates a steering angle to reduce the path tracking error of a vehicle at a point located at a certain `look-ahead distance' on the reference path \cite{snider2009automatic}. We first generated a path around the track and around obstacles that served as a reference for the controller to follow. A look-ahead distance of 3 m was chosen for the controller. Along with the generated reference path, the controller used the longitudinal and lateral coordinates of the vehicle and the heading angle generated by the CarSim vehicle model in real-time to generate the desired steering wheel angle that would achieve path tracking. A controller commanded a torque signal to the motor proportional to the difference between the actual and desired steering wheel angle. Different proportional gains were used for low and high impedance automation systems as further described in the following subsection. A self-aligning (or self-centering) road torque was further added to the automation torque feedback. %The difference between the commanded and actual angles was felt by the driver as torque feedback at the steering wheel and provided haptic cues for automation action. 
%Added to the automation torque feedback was a self-aligning (or self-centering) road torque. %The self-aligning torque arises from the tire-road interaction and is transmitted to the driver through the steering system elements connecting the tires to the steering wheel \cite{bhardwaj2019estimating}. 
The self-aligning torque was designed to be proportional to the steering angle. The proportional gain used in the design was 1.98 N-m/rad. %obstacle avoidance. 
\subsection{Coupling Schemes}

The study employed a between-subject design with one factor (coupling scheme) at three levels. The three coupling schemes between the human driver and the automation system were: \textit{Coupled Low Impedance}, \textit{Coupled  High Impedance}, and \textit{Decoupled with Feedback}. All schemes had haptic feedback during obstacle avoidance. Automation only turned on when an obstacle appeared on the road. The trigger that invoked the automation system is further described in the next subsection.

In the \textit{Decoupled with Feedback} scheme, participants had no control over the vehicle trajectory. Participants could, however, feel the automation torque feedback on the steering wheel. The proportional gain in \textit{Decoupled with Feedback} scheme was 5.96 N-m/rad. Drivers could also move the steering wheel in \textit{Decoupled with Feedback} scheme; however, the driver's steering input was ignored and only the steering angle produced by the automation system was passed to the CarSim model to maneuver the vehicle in the virtual environment.

%could read the steering angle and understand the vehicle trajectory through the visual feedback on the screen and haptic feedback on the steering wheel but could not influence the tire-road angle. 

In the \textit{Coupled Low Impedance} and \textit{Coupled  High Impedance} schemes, drivers could influence the vehicle trajectory by changing the steering angle. Participants could take over control by increasing their grip and imposing a torque on the steering wheel. Conversely, drivers could yield control to the automation system by relaxing their grip (reducing arm impedance) on the steering wheel. In the \textit{Coupled Low Impedance} case, the proportional gain used to determine the automation torque feedback was about three times lower and hence the haptic feedback was weaker than in the \textit{Coupled  High Impedance} case. The proportional gain in the \textit{Coupled Low Impedance} scheme was 5.96 N-m/rad whereas the gain in the \textit{Coupled High Impedance} scheme was 18.46 N-m/rad. As a result, it was also easier to take over control and fight the automation system in the \textit{Coupled Low Impedance} case than it was in the \textit{Coupled  High Impedance} case. Also note that since the proportional gains used in \textit{Coupled Low Impedance} and \textit{Decoupled with Feedback} schemes were the same, the torque feedback experienced in the two schemes was similar. 

\subsection{Experiment Procedure} \label{procedure}
The forty-eight participants recruited to the study were randomly divided into three groups (\textit{Coupled Low Impedance} (CLI), \textit{Coupled High Impedance} (CHI), and \textit{Decoupled with Feedback} (DwF)) of 16 participants each (9 males, 7 females). %The average age and driving experience of participants in the three groups were comparable.

The driving task was to keep the vehicle centered in the right lane of the two-way road and avoid any obstacles that appeared in the lane. To help the driver with lane centering, a lane departure warning appeared on the virtual dashboard (Fig. \ref{1cc}) when the deviation of the vehicle from the center of the right lane exceeded 0.6 m (the lane was 4 m wide).
%\todo[inline, backgroundcolor=blue!10!white,bordercolor=white]{[SP] Consider adding, maybe in parenthesis, the total lane width for context or a brief reasoning for why 0.6 m was chosen.}
Obstacles in the form of pedestrians, deer, or other vehicles unexpectedly entered the road from the right side of the driving lane (Fig. \ref{1c}) and stopped at the center of the lane. Time available to avoid the obstacles was about one second. %\todo[inline, backgroundcolor=blue!10!white,bordercolor=white]{[SP] Define "time headway".}
As soon as the obstacle stopped, the automation system performed an emergency steering intervention towards the left to help the driver avoid the obstacle. During the steering intervention, the lane departure warning disappeared and an `AUTOMATION IS ON' notification appeared on the virtual dashboard to indicate that the automation system was active. %\todo[inline, backgroundcolor=blue!10!white,bordercolor=white]{[SP] "on" or "active"?}
After avoiding the obstacle, the automation system returned the vehicle back to the center of the right lane at which point a take-over-request (TOR) notification `TAKE OVER CONTROL' appeared on the virtual dashboard. Four seconds after the first appearance of the TOR, monotone auditory alerts generating one ``beep" every two seconds were sent from a speaker to remind the driver to take over. The notifications and the auditory alert turned off as soon as the driver pressed the red button, took back control, and resumed manual driving. %\todo[inline, backgroundcolor=blue!10!white,bordercolor=white]{[SP] Do you want to add anything about how this is how Tesla Autopilot warns the driver to apply a steering torque to indicate they are indeed holding onto the handwheel?  Or maybe just keep this in mind for reviewer comments later on why a beep was chosen.}

Each participant was first given instructions on the screen explaining the driving task, the virtual environment, the virtual dashboard, and the coupling scheme that they were assigned. Participants were asked to drive as close as possible to the center of the right lane and look for the lane departure warning. This instruction was given to ensure that all the participants were at the center of their lanes when an obstacle appeared in their lane. Participants were told that the obstacles would appear suddenly and that the automation would always turn on and help them avoid the obstacle. Participants were also advised to keep their hands on the steering wheel when the automation performed an obstacle avoidance maneuver. 

Next, participants completed two 6-minute training trials with one obstacle in each trial and ten formal trials. %Each trial was six minutes long whether it was a testing trial or a training trial.
There was a minute-long break between all trials. The first nine of the ten formal trials were randomized and had eight obstacles in total that were all avoided by the automation system. %There were two training trials with one obstacle in each trial, and ten testing trials with eight obstacles in total that were avoided by the automation system. The first nine out of ten trials were randomized. 
Out of the nine trials, three trials had no obstacles, four trials had one obstacle each, and two trials had two obstacles each. Moreover, trials were designed to have different surroundings (weather and time of day varied between trials) and random start positions and obstacle locations. These measures were taken to discourage any learning and adaptation effects.

\begin{figure}
\centering
\begin{subfigure}[b]{0.4\textwidth}
   \includegraphics[width=1\linewidth]{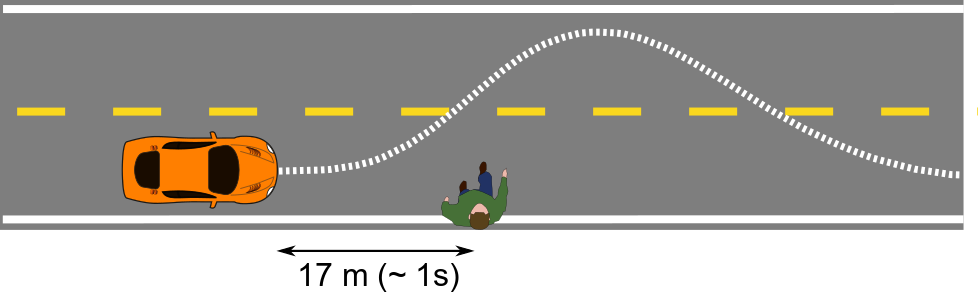}
   \caption{Intended Automation}
   \label{A1} 
\end{subfigure}
\par\medskip
\begin{subfigure}[b]{0.4\textwidth}
   \includegraphics[width=1\linewidth]{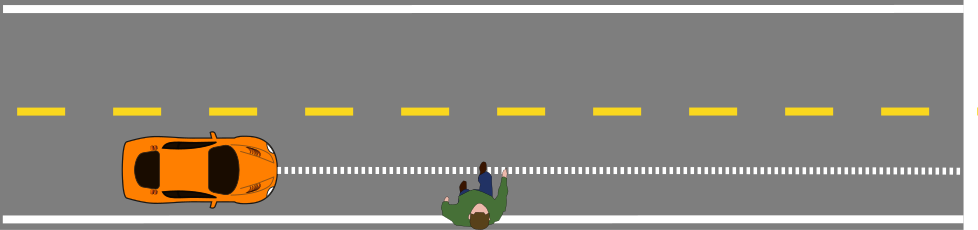}
   \caption{Idle Automation}
   \label{A2}
\end{subfigure}
\par\medskip
\begin{subfigure}[b]{0.4\textwidth}
   \includegraphics[width=1\linewidth]{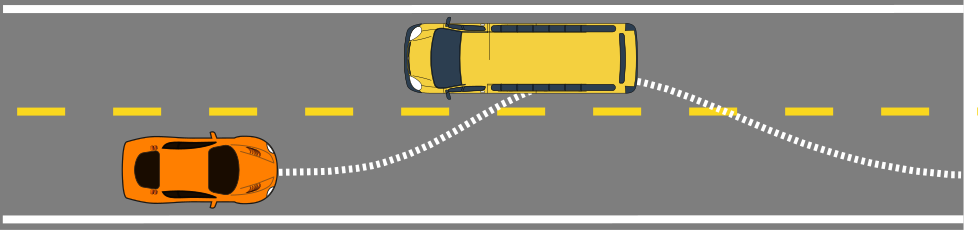}
   \caption{Adversarial Automation}
   \label{A3}
\end{subfigure}
\caption[]{Three types of automation behaviors designed and tested in the experiment.}\label{fig2}
\end{figure}

The eight obstacles in the first nine formal trials were all avoided by the automation which worked as intended (see  Fig. \ref{A1}). This resulted in a total of 128 ``intended" automation obstacles for each of the three groups. The tenth trial always involved an unexpected automation failure, either idle automation (automation failed to activate in the presence of an obstacle) or adversarial automation (automation initiated a maneuver into oncoming traffic in the absence of an obstacle), as shown in Fig. \ref{A2} and Fig. \ref{A3}. Half of the participants in each coupling scheme experienced idle automation in their last trial and the other half experienced adversarial automation. This resulted in a total of eight idle automation obstacles and eight adversarial automation obstacles in each group. At the end of the experiment, participants were asked to fill out a debriefing questionnaire that was used to gather participant feedback on the three automation schemes.

%The eight obstacles in the first nine testing trials were all avoided by the automation system. This automation system was called \emph{intended automation} as shown in Fig. \ref{A1}. For 16 participants in each scheme, this resulted in a total of 128 intended automation obstacles for each scheme. The last testing trial always had an unexpected automation failure. There were two types of failed automation systems: idle automation and adversarial automation as shown in Fig. \ref{A2} and Fig. \ref{A3}. In the idle automation case, automation did not drive around the obstacle and the driver was left in control, whereas in the adversarial automation case the automation system unnecessarily turned on and drove through a bus parked on the other side of the road. Half of the participants in each coupling scheme experienced idle automation in their last trial and the other half experienced adversarial automation. This resulted in a total of eight idle automation obstacles and eight adversarial automation obstacles for each scheme. At the end of the experiment, participants were asked to fill out a debriefing questionnaire that was used to gather participant feedback on the three schemes.

\subsection{Performance Metrics and Data Analysis}

\begin{figure}[h!]
\begin{center}
\includegraphics[width=0.45\textwidth]{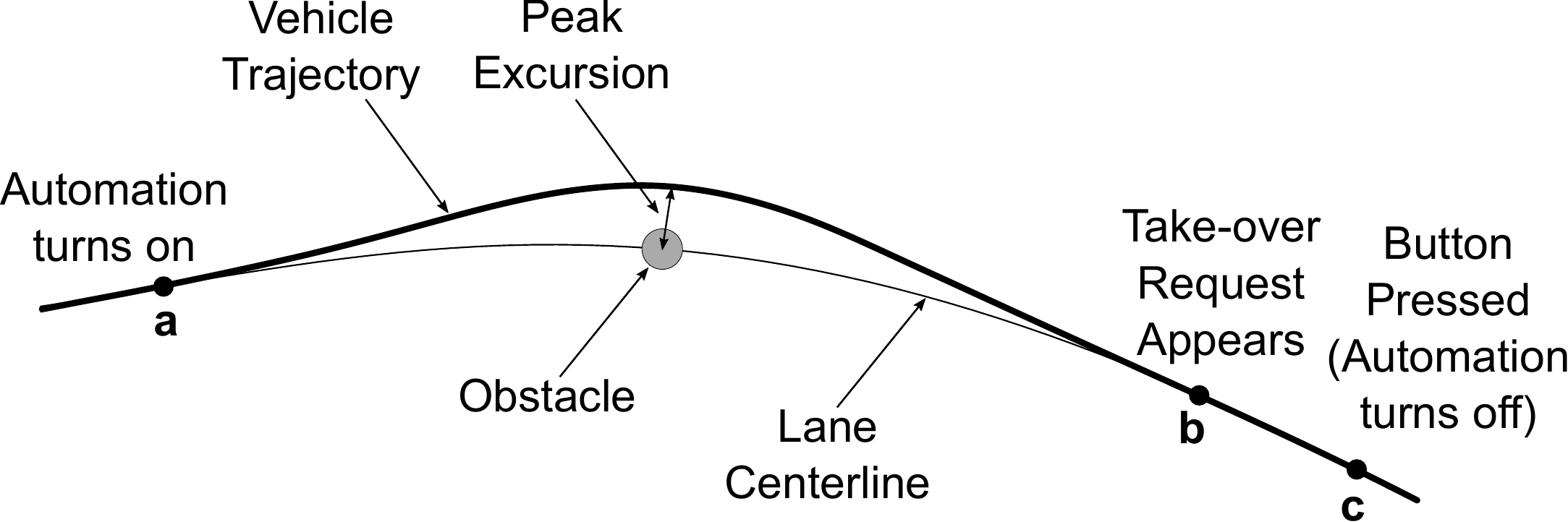}
\caption{A sample trajectory around the obstacle. The automation turned on at point \textbf{a} which was located 17 m (about 1 second) before the obstacle. A take-over-request appeared on the screen at point \textbf{b} when the automation brought the vehicle back to the center of the right lane. The automation turned off at point \textbf{c} when the subject pressed the red button. Maximum deviation from the center of the right lane was defined as the Peak Excursion. The time taken by the vehicle to travel from \textbf{a} to \textbf{b} was defined as the Excursion Time, and the time taken to travel from \textbf{b} to \textbf{c} was defined as the Handback Time.} \label{metrics}
\end{center}
\end{figure}

There were four dependent measures in this study: (1) Obstacle Hits, (2) Peak Excursion, (3) Excursion Time, and (4) Handback Time. The first metric, Obstacle Hits, was defined as the total number of collisions with obstacles and was analyzed separately for intended automation, idle automation, and adversarial automation. The remaining three metrics were analyzed only for the intended automation condition. Peak Excursion was measured as the absolute maximum lateral deviation of the vehicle away from the center of the driving lane while avoiding the obstacle (as shown in Fig. \ref{metrics}). Excursion Time was defined as the time between the instant the vehicle first departed the lane to avoid the obstacle and the instant the vehicle came back to the center of the driving lane. Finally, Handback Time was the time taken by the driver to press the red button to turn the automation off after the TOR appeared for the first time on the screen. 

%Four metrics were defined in this study to quantify driving performance and enable comparison between the three coupling schemes. The first metric, Obstacle Hits, was the number of total hits in each control scheme. This metric was separately analyzed for the three types of obstacles encountered: obstacle hits for intended automation, for idle automation, and for adversarial automation. The rest of the metrics were only analyzed for the intended automation obstacles. The second metric, Peak Excursion, was the absolute maximum lateral deviation of the vehicle away from the center of the right lane while avoiding the obstacle (as shown in Fig. \ref{metrics}). The third metric, Excursion Time, was defined as the time between the instances when the vehicle first departed the lane to avoid the obstacle to the instant when the vehicle came back to the center of the right lane. Finally, the Handback Time was the time taken by the driver to press the button to turn the automation off after the TOR appeared for the first time on the screen. Note that  Peak Excursion, Excursion Time, and Handback Time were only computed around the obstacles that were not hit. %\todo[inline, backgroundcolor=blue!10!white,bordercolor=white]{[SP] Make the capitalization and hyphens of the terms "Take-over Time" and "take-over-request" consistent?}

Obstacle Hits were analyzed using binary logistic regression. All other metrics were analyzed using linear mixed models with coupling scheme as a fixed factor and subject ID as a random factor. The significance level was set at $p<0.05$. Post-hoc Bonferroni tests were conducted to perform pairwise comparisons between the three coupling schemes.

%In this study, there was only one factor (coupling scheme) with three levels. The four performance metrics served as the dependent variables. %Data analysis was performed in IBM SPSS Statistics version 26.
%To analyze the main effect of the coupling scheme, the Obstacle Hits metric was analyzed using binary logistic regression whereas all the other metrics were analyzed using linear mixed models. Coupling scheme was chosen as a fixed factor and subject ID was chosen as a random factor. A $p$-value of 0.05 was set to determine significance. Post-hoc Bonferroni tests were conducted to perform pairwise comparisons between the three coupling schemes.

\begin{figure*}[h!]
\centering
   \includegraphics[width=0.8\linewidth]{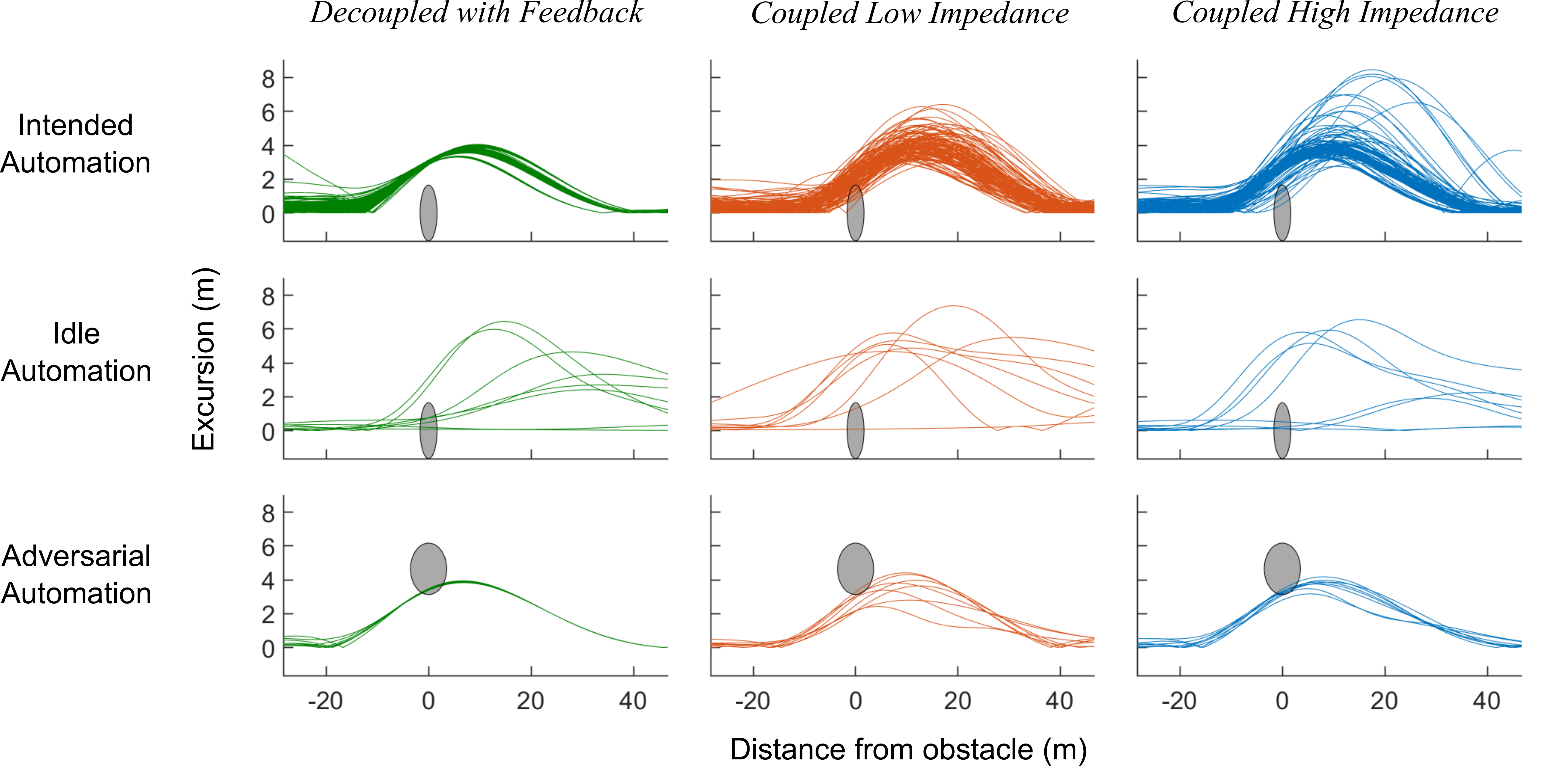}
\caption{Driving trajectories around obstacles for all 48 participants, separated by coupling scheme and automation behavior. Obstacles are depicted to scale by grey ellipses in each plot. Intersection of trajectories with the obstacles indicate obstacle hits. (Note that the obstacle in adversarial automation case had a different size as shown in Fig. \ref{fig2}.)}
\label{trajs}
\end{figure*}

\section{Results}

Differences between the three coupling schemes were apparent in the Obstacle Hits, Peak Excursion, and Excursion Time metrics. Fig. \ref{trajs} shows the driving trajectories taken around the obstacles by all participants in each coupling scheme. The approximate obstacle size is indicated by a grey ellipse and any intersection of the trajectories with the ellipse denotes an obstacle hit. More intended automation obstacles were hit in the \textit{Coupled Low Impedance} group than in the \textit{Coupled High Impedance} and the \textit{Decoupled with Feedback} groups. In contrast, fewer idle and adversarial automation obstacles were hit in the \textit{Coupled Low Impedance} group, compared to the other two groups. For the intended automation obstacles, \textit{Coupled High Impedance} resulted in large excursions around the obstacles whereas the \textit{Decoupled with Feedback} scheme resulted in the smallest excursions.

%Differences between the three coupling schemes were apparent in the obstacle hits and excursion metrics. Fig. \ref{trajs} shows the driving trajectories taken around the obstacles by all participants in each coupling scheme. The obstacles are shown by grey ellipses and any intersection of the trajectories with the ellipse denotes an obstacle hit. A larger number of intended automation obstacles were hit in the \textit{Coupled Low Impedance} scheme than in the \textit{Coupled High Impedance} and the \textit{Decoupled with Feedback} schemes. On the contrary, a fewer number of idle and adversarial automation obstacles were hit in the \textit{Coupled Low Impedance} scheme than the other two coupling schemes. For the intended automation case, \textit{Coupled High Impedance} resulted in large excursions around the obstacles whereas the \textit{Decoupled with Feedback} scheme resulted in the smallest excursions.

\subsection{Obstacle Hits}

\begin{figure}[h!]
\begin{center}
\includegraphics[width=0.49\textwidth]{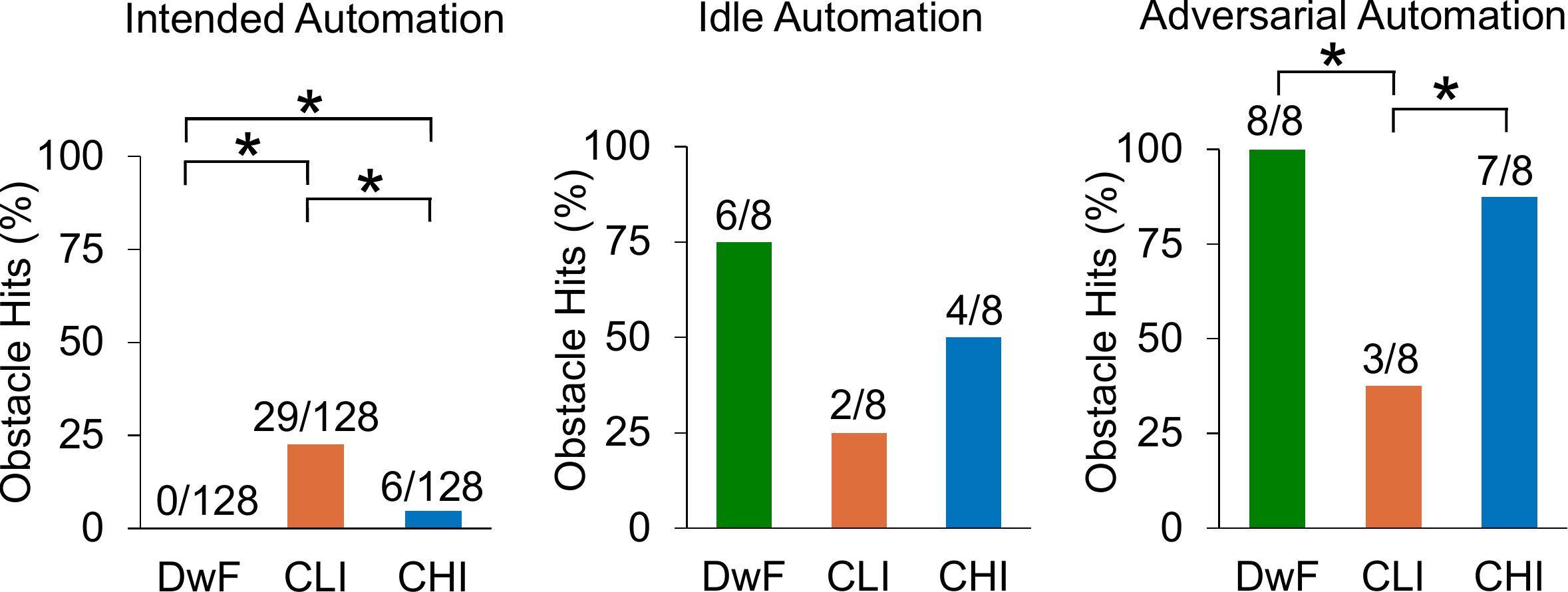}
\caption{Percent obstacle hits for three types of obstacles separated by coupling scheme. Figures on the top of each bar (X/Y) indicate the number of hits (X) out of the number of obstacles encountered (Y) in each case.} \label{R1}
\end{center}
\end{figure}

\subsubsection{Intended Automation}
Out of the 128 intended automation obstacles, the \textit{Decoupled with Feedback} group had no obstacle collisions (Fig. \ref{R1}). On the other hand, the \textit{Coupled  High Impedance} group had six and the \textit{Coupled Low Impedance} group had 29 collisions. 

Analysis on the Obstacle Hits metric indicated a main effect of coupling scheme ($F(2,381) = 4.791,\: p=0.009$). Post-hoc comparisons further revealed that the likelihood of a hit for the \textit{Coupled Low Impedance} group was significantly higher than for the \textit{Coupled  High Impedance} group ($p=0.006$) and the \textit{Decoupled with Feedback} group ($p<0.001$). Moreover, the likelihood of a hit in the \textit{Coupled  High Impedance} group was significantly higher than the \textit{Decoupled with Feedback} group ($p=0.033$).

%the \textit{Coupled  High Impedance} scheme ($\chi^2(1) = 18.158, p<0.001$) and the \textit{Decoupled with Feedback} scheme ($\chi^2(1) = 9.1, p=0.003$). Moreover, likelihood of a hit in the \textit{Coupled  High Impedance} scheme was significantly higher than the \textit{Decoupled with Feedback} scheme ($\chi^2(1) = 6.232, p=0.001$).

\subsubsection{Idle Automation}
In the Idle Automation case, the \textit{Coupled Low Impedance} group had only two hits out of eight obstacles. Six out of eight obstacles were hit in the \textit{Decoupled with Feedback} scheme, while four out of eight obstacles were hit in the \textit{Coupled High Impedance} group (Fig. \ref{R1}). The effect of coupling scheme on hits was not significant ($p=0.147$).

\subsubsection{Adversarial Automation}
Out of eight adversarial automation obstacles, \textit{Coupled  High Impedance} group had seven hits, \textit{Decoupled with Feedback} group had eight hits, and \textit{Coupled Low Impedance} group had three hits (see Fig. \ref{R1}).
%In the adversarial automation case also, the \textit{Coupled Low Impedance} group had fewest hits. Three out of eight obstacles were hit, were the fewest: three out of eight adversarial automation obstacles were hit in the \textit{Coupled Low Impedance} scheme. All eight obstacles were hit in the \textit{Decoupled with Feedback} scheme, whereas seven out of eight obstacles were hit in the \textit{Coupled  High Impedance} scheme (Fig. \ref{trajs2}a). 
There was a significant effect of coupling scheme in the adversarial automation case ($F(2,21) = 6.682,\: p=0.006$). Post-hoc Bonferroni tests revealed that the likelihood of a hit for the \textit{Coupled Low Impedance} group was significantly lower than for the \textit{Decoupled with Feedback} group ($p=0.007$) and the \textit{Coupled  High Impedance} group ($p=0.035$). No significant difference was found between the \textit{Decoupled with Feedback} and the \textit{Coupled  High Impedance} groups.
%As expected, the \textit{Decoupling with Feedback} scheme had no obstacle collisions because neither the automation action nor the vehicle trajectory were influenced by driver input.

\subsection{Peak Excursion}

\begin{figure}[h!]
\centering
\begin{subfigure}[b]{0.16\textwidth}
  \includegraphics[width=1\linewidth]{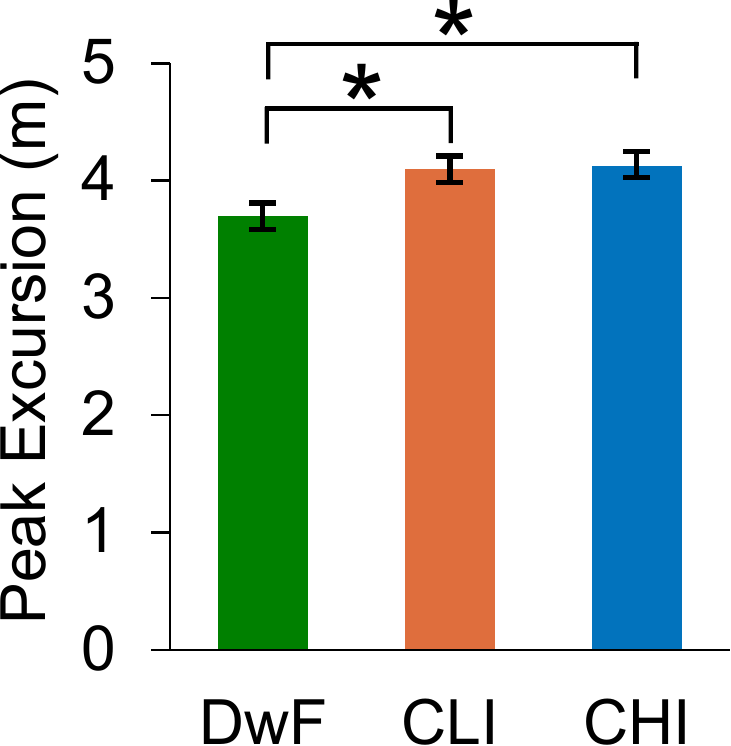}
  \caption{Peak Excursion}
  \label{R2} 
\end{subfigure}\hspace{3em}
\begin{subfigure}[b]{0.16\textwidth}
  \includegraphics[width=1\linewidth]{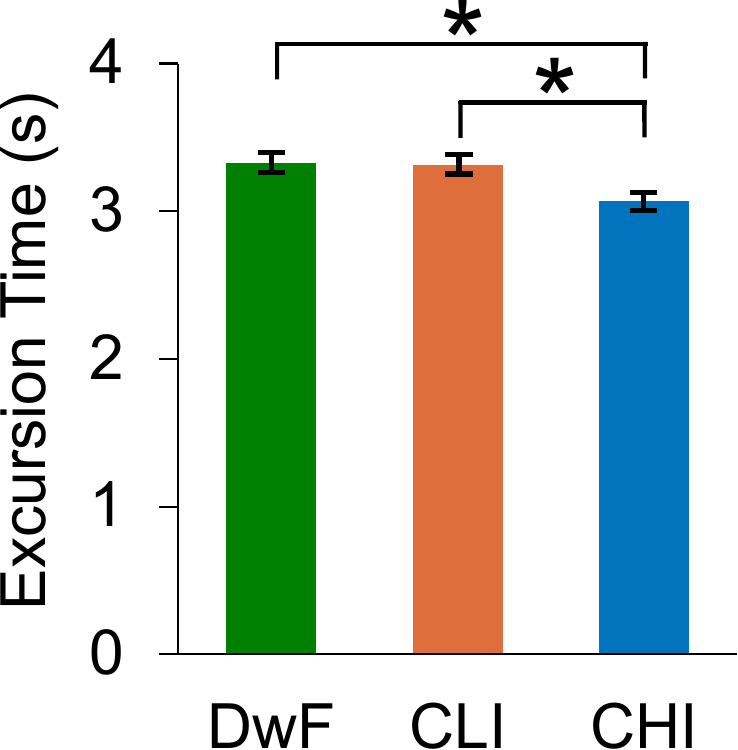}
  \caption{Excursion Time}
  \label{R3} 
\end{subfigure}
\caption{(a) Mean Peak Excursion and (b) Mean Excursion Time for the three coupling schemes. Error bars indicate $\pm$1 standard error of mean.}
\label{trajs2}
\end{figure}%The asterisks indicate a significant different between the two coupling schemes.

Peak Excursion was used to gauge which coupling scheme produced the largest deviations from the center of the driving lane. %Fig. \ref{trajs} shows the driving trajectories taken around the obstacles by all the participants in each coupling scheme. The obstacles are denoted by a black ellipse and the y-axis denotes the lateral deviation. Any intersections of the trajectory with the ellipse denote a collision. The maximum y-axis value for each trajectory is the Peak Excursion. %Clearly, the mean and standard errors of Peak Excursion was the largest for \textit{Coupled  High Impedance} scheme and lowest in the \textit{Decoupled with Feedback} scheme.Larger deviations from lane center of Peak Excursion indicate inefficient and potentially dangerous maneuvers. 
%The bar chart in Fig. \ref{R2} shows the mean and standard error for the Peak Excursion metric. 
Peak Excursion differed significantly between the three coupling schemes ($F(2,346) = 4.413,\: p=0.013$) (see Fig. \ref{R2}). Post-hoc Bonferroni tests revealed that the \textit{Decoupled with Feedback} group had a significantly lower mean Peak Excursion than the \textit{Coupled  High Impedance} group (3.70 m vs. 4.12 m, $p=0.025$) and the \textit{Coupled Low Impedance} group (3.70 m vs. 4.08 m, $p=0.029$). No other significant differences were found. 

\subsection{Excursion Time}

Excursion Time indicated how much time was spent away from the lane center during obstacle avoidance. There was a main effect of coupling scheme on Excursion Time 
%Excursion Time was used to gauge which coupling scheme resulted in more time spent away from the lane center during obstacle avoidance. The bar chart in Fig. \ref{R3} show the mean and standard errors for Excursion Time. Excursion Time was also found to be significantly different between the three coupling schemes
($F(2,346) = 4.413,\: p=0.003$) (see Fig. \ref{R3}). Post-hoc tests showed that the \textit{Coupled  High Impedance} scheme had significantly lower mean Excursion Time than both the \textit{Decoupled with Feedback} scheme (3.07 s vs. 3.33 s, $p=0.005$) and the \textit{Coupled Low Impedance} scheme (3.07 s vs. 3.32 s, $p=0.01$). No other significant differences were found.

\subsection{Handback Time}

Handback Time was used to measure which coupling scheme encouraged faster automation-to-driver transitions. %The mean Handback Time for the \textit{Decoupled with Feedback} scheme was the lowest whereas for the \textit{Coupled Low Impedance} scheme it was the highest. However, 
%Analysis on the Handback Time metric revealed the differences between the mean Handback Time for the three schemes were not statistically significant 
There were no significant differences between the mean Handback Time for the three groups ($p=0.348$).

\section{Discussion}
In this driving simulator study, we compared the obstacle avoidance performance for three automatic steering intervention schemes which differed in the amount of control authority provided to the automation system. In the \textit{Decoupled with Feedback} scheme, the driver and the steering wheel were decoupled from the tires, and automation had full control over the vehicle. In the \textit{Coupled High Impedance} and \textit{Coupled Low Impedance} schemes, the steering wheel was coupled to the tires, and the automation system was provided high control authority or low control authority, respectively. When working properly, the automation helped the driver avoid an obstacle that appeared unexpectedly on the road (intended automation). One of two types of automation failure was simulated during the last trial: the automation would either fail to activate when an obstacle appeared (idle automation) or it would initiate a maneuver into oncoming traffic in the absence of an obstacle (adversarial automation).

The likelihood of a collision when the automation worked properly was significantly lower in the \textit{Coupled High Impedance} group than in the \textit{Coupled Low Impedance} group. During emergency situations, drivers tend to fight any steering intervention and act as a disturbance to the automation \cite{seto2004research, heesen2014interaction}. Hence, a lower likelihood of collision in the \textit{Coupled High Impedance} group suggests that a high impedance automation system has the potential to improve safety by reducing the influence of driver disturbance during emergency situations (as also hypothesized in \cite{heesen2014interaction, sieber2015automatic}). However, the \textit{Coupled High Impedance} group still had a significantly higher likelihood of a collision than the \textit{Decoupled with Feedback} group implying that allowing even a minor driver disturbance at the steering wheel imposes some risk of collision. 

A different picture emerges however in the case of automation faults. Here, the likelihood of a collision did not differ significantly between the \textit{Coupled High Impedance} and the \textit{Decoupled with Feedback} groups. This suggests that during automation faults a high impedance automation might behave similarly to an automation system in which the driver is decoupled from the tires. Especially when the automation initiated a maneuver in the absence of an obstacle (adversarial automation), participants in the \textit{Coupled High Impedance} group reported that they recognized the automation failure but found it difficult to override the automation in a timely fashion to prevent the collision. This may be the result of startle and confusion on the part of the driver who, despite being aware of the option to override the automation, was not able to exert control quickly and decisively. In the presence of an obstacle, a driver is likely to monitor the automation to ensure that it takes action. This mental readiness allows for faster and more effective interventions. In contrast, in the absence of an obstacle, the driver experiences a surprise, resulting in a longer response time. A longer response time in combination with larger muscular effort required to overpower a high impedance automation significantly increases the chances of a collision. %basically renders the coupling between human and automation ineffective.
On the other hand, the likelihood of a collision in the \textit{Coupled Low Impedance} group was significantly lower than for the other two coupling schemes. The \textit{Coupled Low Impedance} scheme enables the driver to `edit' automation control inputs fairly quickly and easily when necessary. These results are consistent with the lumberjack analogy \cite{sebok2017implementing}; when automation worked properly, the increased degree of automation could improve performance, but in the failure conditions, higher levels of automation led to more significantly degraded performance.%[I think we need to include more references to earlier work in the Discussion to compare and contrast our findings with theirs]

% new paragraph on idle automation: 
In the idle automation case, we had expected that the \textit{Decoupled with Feedback} group would promote driver complacency and over-reliance on the automation's action \pcite{young2007driving, endsley1995out}, and would therefore result in significantly more hits than the \textit{Coupled High Impedance} and \textit{Coupled Low Impedance} groups. The participants in the \textit{Decoupled with Feedback} group did report that they relied on automation to avoid the obstacle and either did not react or reacted very late to the automation dropout (see also Fig. \ref{trajs}). However, no statistically significant differences in the number of hits were seen between the three coupling conditions in the idle automation case. %Perhaps more data is required to investigate the effect of decoupling during automation dropouts. %The feedback obtained from the participants merits further investigation with perhaps more data and trajectory analysis. %and perhaps more data should be collected before inferring that a decoupled steering wheel in fact results in over-reliance on automation.

The excursions around obstacles also reveal some important differences in driver behavior for the three coupling schemes. Not surprisingly, the excursions were significantly larger in the coupled driving schemes, compared to the decoupled scheme (Fig. \ref{R2}). This can be attributed to the added driver input in the coupled driving schemes. More importantly, in the \textit{Coupled High Impedance} scheme, a few trajectories exhibited overshoots beyond the edge of the road (see Fig. \ref{trajs}). Consistent with the findings in \cite{petermeijer2015should} and \cite{mars2014analysis}, some participants in the \textit{Coupled High Impedance} group reported that the transition from manual to automated driving was sudden and discomforting, and that the automation acted too strongly and aggressively. As a result, the participants may have initially fought the automation but after seeing the obstacle perhaps relinquished control to automation resulting in the overshoots.

The excursions in the \textit{Coupled High Impedance} group were significantly shorter than in the other two groups. One possible explanation for this finding is that participants were uncomfortable with the rather powerful automation that they found difficult to override, and therefore wanted to return to the center of the right lane as soon as possible to take back control of the vehicle. Finally, we expected that participants would take longer to take back control from the automation in the \textit{Decoupled with Feedback} group because of being out-of-the-loop with a decoupled steering wheel during obstacle avoidance \cite{endsley1995out,  fricke2015driver}. However, the differences in Handback Time between the groups were not significant.

In summary, the results of this study highlight a trade-off in automation design for emergency situations: high impedance automation can significantly reduce unwarranted driver input on the steering wheel during emergency situations but may cause driver discomfort and may be too strong to override during automation faults. This result is consistent with the hypotheses and findings presented in the past \cite{parasuraman2000model, sebok2017implementing, fricke2015driver, sieber2015automatic, petermeijer2015should, mars2014analysis, zwaan2019haptic}. Contrary to expectations, decoupling the driver during emergency interventions did not significantly increase the time required for the driver to resume control or the number of collisions during automation dropouts. To combine the advantages of low and high impedance automation, an adaptive impedance system could be designed that would assume a high level of authority during emergency situations in which the automation has high confidence, and a low level of authority during situations in which the automation has low confidence to give override power to the human \cite{9147984, zwaan2019haptic}. The design challenge for such an automation system would be to modulate automation impedance as a function of driver intention, sensor precision, and environmental complexity.

\bibliography{SMC.bib}
\bibliographystyle{ieeetr}

\end{document}